\documentclass[12pt]{article}
\usepackage{geometry}
\usepackage{amsmath, amssymb, graphicx,fullpage,color,mathtools,amsthm,xcolor}
\usepackage{caption, subcaption}
\usepackage{algorithm, algorithmic}
\usepackage{multirow}

\usepackage[normalem]{ulem}

\usepackage{microtype}
\usepackage{hyperref,color}

\definecolor{webgreen}{rgb}{0,.35,0}
\definecolor{webbrown}{rgb}{.6,0,0}
\definecolor{RoyalBlue}{rgb}{0,0,0.9}
\definecolor{purp}{rgb}{0.6,0.05,0.8}
\definecolor{ora}{rgb}{0.7,0.35,0.02}

\hypersetup{
   colorlinks=true, linktocpage=true, 
   urlcolor=webbrown, linkcolor=RoyalBlue, citecolor=black,
   pdfauthor={Gary P. T. Choi, Mahmoud Shaqfa},
   pdfsubject={Hemispheroidal parameterization and harmonic decomposition}
}

\begin{document}

\author{Gary P. T. Choi\href{https://orcid.org/0000-0001-5407-9111}{\protect\includegraphics[scale=.050]{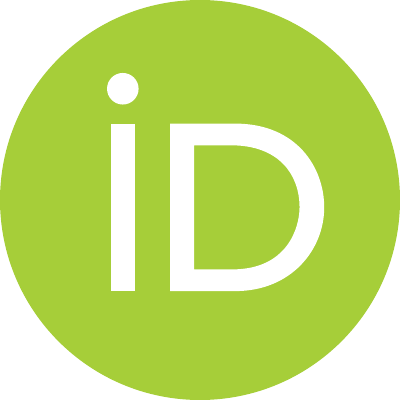}}$^{1,\ast}$, Mahmoud Shaqfa\href{https://orcid.org/0000-0002-0136-2391}{\protect\includegraphics[scale=.050]{ORCID.pdf}}$^{2}$\\
\\
\footnotesize{$^{1}$Department of Mathematics, The Chinese University of Hong Kong, Hong Kong}\\
\footnotesize{$^{2}$Department of Mechanical Engineering, Massachusetts Institute of Technology, Cambridge, MA, USA}\\
\footnotesize{$^\ast$To whom correspondence should be addressed; E-mail: ptchoi@cuhk.edu.hk}
}
\title{Hemispheroidal parameterization and harmonic decomposition of simply connected open surfaces}

\date{}
\maketitle

\begin{abstract}
Spectral analysis of open surfaces is gaining momentum for studying surface morphology in engineering, computer graphics, and medical domains. This analysis is enabled using proper parameterization approaches on the target analysis domain. In this paper, we propose the usage of customizable parameterization coordinates that allow mapping open surfaces into oblate or prolate hemispheroidal surfaces. For this, we proposed the usage of Tutte, conformal, area-preserving, and balanced mappings for parameterizing any given simply connected open surface onto an optimal hemispheroid. The hemispheroidal harmonic bases were introduced to spectrally expand these parametric surfaces by generalizing the known hemispherical ones. This approach uses the radius of the hemispheroid as a degree of freedom to control the size of the parameterization domain of the open surfaces while providing numerically stable basis functions. Several open surfaces have been tested using different mapping combinations. We also propose optimization-based mappings to serve various applications on the reconstruction problem. Altogether, our work provides an effective way to represent and analyze simply connected open surfaces.
\end{abstract}

\section{Introduction} \label{sect:introduction}
In many scientific and engineering problems, a fundamental task is to perform shape analysis of a given surface. To achieve this, one common approach is to parameterize the surface onto a standardized domain and then utilize certain functions defined on the standardized domain to represent the shape effectively. Over the past several decades, numerous works on surface parameterization~\cite{floater2005surface, sheffer2007mesh, choi2022recent} and surface harmonics~\cite{hobson1931theory, byerly1959elementary, li1998computations, muller2006spherical, dassios2012ellipsoidal} have been developed. For instance, Brechb{\"u}hler et~al.~\cite{brechbuhler1995parametrization} combined spherical parameterization and spherical harmonics for 3D closed surface description. Gu et~al.~\cite{gu2004genus} developed a harmonic energy minimization method for spherical conformal parameterization and combined it with spherical harmonics. Several other spherical parameterization methods have also been proposed for genus-0 closed surfaces~\cite{haker2000conformal, lai2014folding, choi2015flash, wang2018novel}. Besides spherical parameterization, methods for spheroidal mapping and harmonics~\cite{shaqfa2024spheroid} and ellipsoidal conformal and quasi-conformal parameterization~\cite{choi2024fast} have also been recently developed for genus-0 closed surfaces. 
\par
For simply connected open surfaces, one common approach is to parameterize them onto a planar domain. In particular, existing methods primarily focus on disk conformal parameterization~\cite{choi2015fast, yueh2017efficient, choi2018linear}, disk area-preserving parameterization~\cite{choi2018density, yueh2019novel, lyu2024bijective}, and parameterization onto other predetermined~\cite{choi2020area} or flexible~\cite{levy2002least, mullen2008spectral, yang2008optimal} planar domains. By combining disk parameterization and disk harmonics, Shaqfa et~al.~\cite{shaqfa2023disk} developed a framework for analyzing curved and nominally flat rough surfaces. Besides, hemispherical parameterization~\cite{giri2021open} and hemispherical harmonics~\cite{huang2006hemispherical} were used for medical shape analysis. Recently, generalizations of them to spherical cap domains have also been proposed~\cite{shaqfa2021spherical, choi2022adaptive} and applied to the analysis of rough interfaces and medical shapes. However, as illustrated by these works, the accuracy and effectiveness of the parameterization and harmonics representation of simply connected open surfaces highly depend on the surface geometry. For instance, for nominally flat open surfaces, using a hemispherical domain for the parameterization may lead to a large geometric distortion.
\par
From the perspective of surface harmonics, the choice of the parameterization domain is constrained by the availability of a separable solution to Laplace's equation. Mathematically, there are $11$ curvilinear coordinate systems where Laplace and Helmholtz equations are separable, e.g., rectangular, spherical, spheroidal, ellipsoidal, cylindrical, and toroidal, to mention but a few, \cite{Morse_Feshbach_1953}. In the case of the spherical cap harmonics (SCH), the basis functions are dependent on the choice of the boundary condition (Neumann and Dirichlet). As expressed by \cite{Haines1985}, the associated Legendre functions, in the spherical cap harmonics, can be written in terms of the hypergeometric functions. Such formulation is not stable for large degrees rather being very slow as shown by \cite{shaqfa2021spherical}. Thus, the need for customizable cap-like domains with efficient harmonic bases persists. 
\par
In this work, we propose a novel framework for the hemispheroidal parameterization and hemispheroidal harmonics decomposition of simply connected open surfaces. More specifically, we first develop efficient methods for parameterizing simply connected open surfaces onto a hemispheroidal shape, covering Tutte parameterizations, angle-preserving parameterizations, area-preserving parameterizations, and parameterizations balancing the angle and area distortions. We then combine the hemispheroidal parameterizations with the hemispheroidal harmonics, which provide unconditionally stable harmonics, for surface representation and analysis. We can obtain the basis for the hemispheroidal harmonics by shifting the associated Legendre polynomials with an affine transformation while preserving the domain's orthonormality. We demonstrate the effectiveness of our framework using a large variety of simply connected open surfaces, including human faces, anatomical structures, and other complex digital surfaces.
\par
The rest of the paper is organized as follows. In Section~\ref{sect:parameterization}, we introduce our proposed methods for hemispheroidal parameterization. In Section~\ref{sect:harmonics}, we describe the computation of hemispheroidal harmonics in detail. In Section~\ref{sect:experiments}, we present experimental results on hemispheroidal parameterization and harmonics using different surfaces. We conclude our work and discuss future directions in Section~\ref{sect:conclusion}.

\section{Hemispheroidal parameterization} \label{sect:parameterization}
Consider a simply connected open surface $\mathcal{M} = (\mathcal{V},\mathcal{E},\mathcal{F})$, where $\mathcal{V}$ is the set of vertices, $\mathcal{E}$ is the set of edges, and $\mathcal{F}$ is the set of triangular faces. The Northern hemispheroidal surface can be defined as a surface of revolution of the confocal elliptic coordinate system, and its implicit form can be written as:
\begin{equation} \label{eqt:hemispheroid}
    \mathcal{H} = \left\{(x,y,z) \in \mathbb{R}^3: \frac{x^2 + y^2}{a^2} + \frac{z^2}{c^2} = 1, \text{with} \quad z \geq 0 \right\},
\end{equation}
where $a,c$ are positive scalars for the semiaxes (radii) of $\mathcal{H}$. Our goal is to compute an optimal hemispheroidal parameterization $f: \mathcal{M} \to \mathcal{H}$ based on certain geometric distortion criteria.

\subsection{Surface registration and sizing the target hemispheroidal domain}
\label{subsec:surface_registraction}
\par
Prior to the mapping process we need to place our open surfaces $\mathcal{M}$ into canonical coordinates. The mesh placement in this reference coordinate is important for a successful parameterization process. We start this registration process by shifting the geometric centroid of the surface to coincide with the origin of the Cartesian coordinates $\in \mathbb{R}^3$. Next, we align the surface boundary $\partial \mathcal{M}$ such that it is parallel to the XY plane. One way to do this is by first fitting a plane to the boundary vertices of the open surface with its current configuration. Afterward, we rotate the whole mesh such that the fitted plane is completely flat and parallel to the XY plane. Such registration processes can be reversed after the final reconstruction of the input surface using the hemispheroidal harmonics (HSOH).
\par
\begin{figure}[t]
    \centering
    \includegraphics[width=0.8\textwidth]{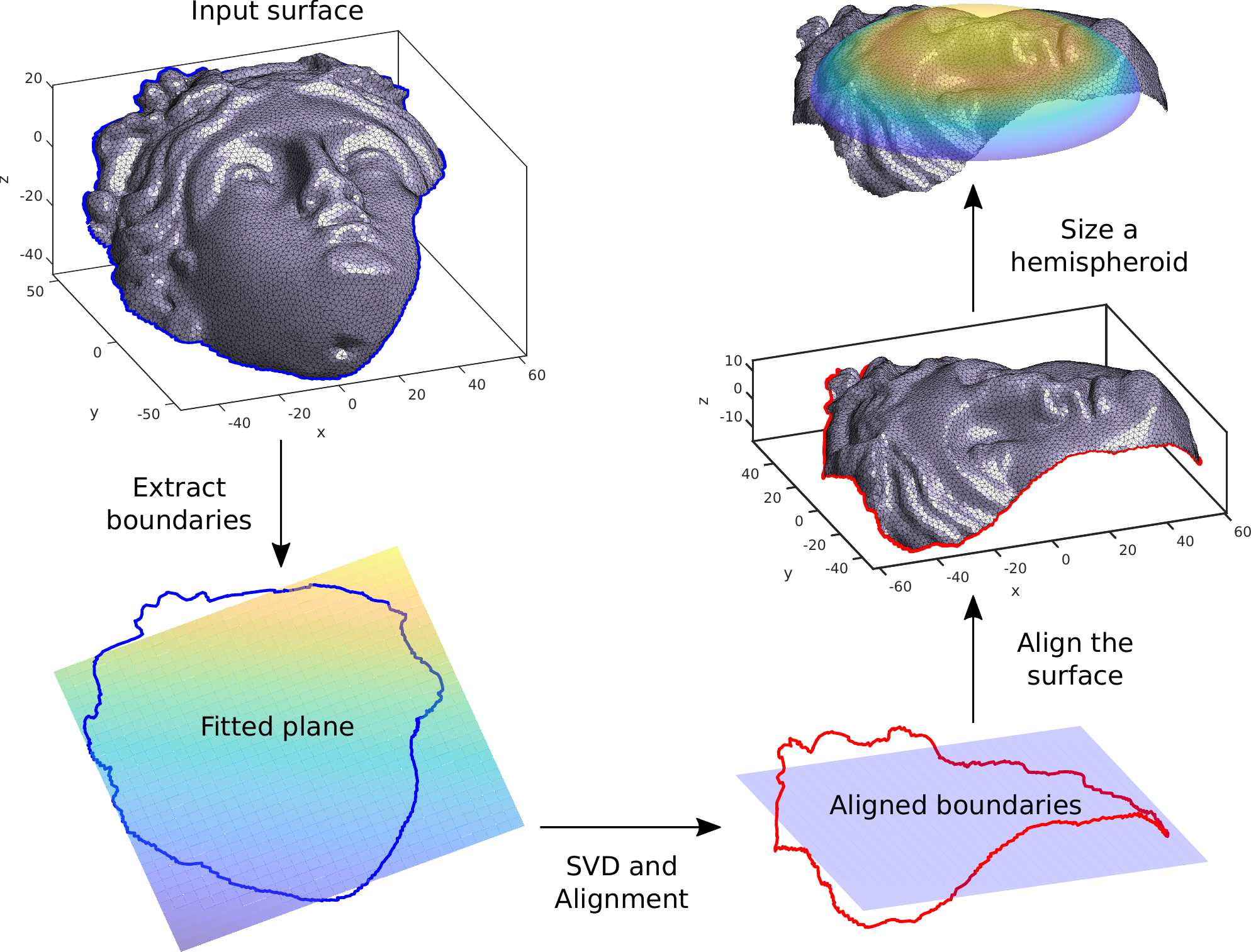}
    \caption{Schematic of the surface alignment process of a given input surface as part of the registration. We start by fitting a plane to the boundary points. Then, we align the points via the SVD algorithm and accordingly align the whole surface, to finally go to the sizing process of a hemispheroid (find the radius $c$ given that $a = 1$).}
    \label{fig:surface_registration}
\end{figure}

The rationale behind using hemispheroidal parameterization is the extra degree of freedom (DOF) we gain for sizing the analysis domains. In our paper, this DOF can be seen as the radius of the hemispheroid resulting in either an oblate or prolate hemispheroidal domain. In our recent work, Shaqfa and van Rees \cite{shaqfa2024spheroid}, we found that fitted spheroidal domains, in the least-squares sense, result in excellent reconstruction results. Though not always optimal, that approach is rather efficient and computationally affordable. Building on that, using the normalized height of the bounding box (BB) for the radius $c$ of the registered surfaces can effectively lead to solutions with excellent reconstruction results. By the normalized height $c$ we mean that the obtained hemispheroid is isotropically normalized such that $a = 1$. Figure \ref{fig:surface_registration} summarizes the surface registration process of a given simply connected open surface.
\par
After obtaining the radius $c$ of the hemispheroidal domain, we can focus on developing hemispheroidal parameterization methods to map $\mathcal{M}$ to the target hemispheroid $\mathcal{H}$. In the following subsection, we introduce three hemispheroidal parameterization methods that start with this registration and sizing process. Later, we use an optimization-based approach to determine the optimal radius $c$ to study the impact of this choice on the overall distortion metrics and reconstruction accuracy.

\begin{figure}[t]
    \centering
    \includegraphics[width=0.95\textwidth]{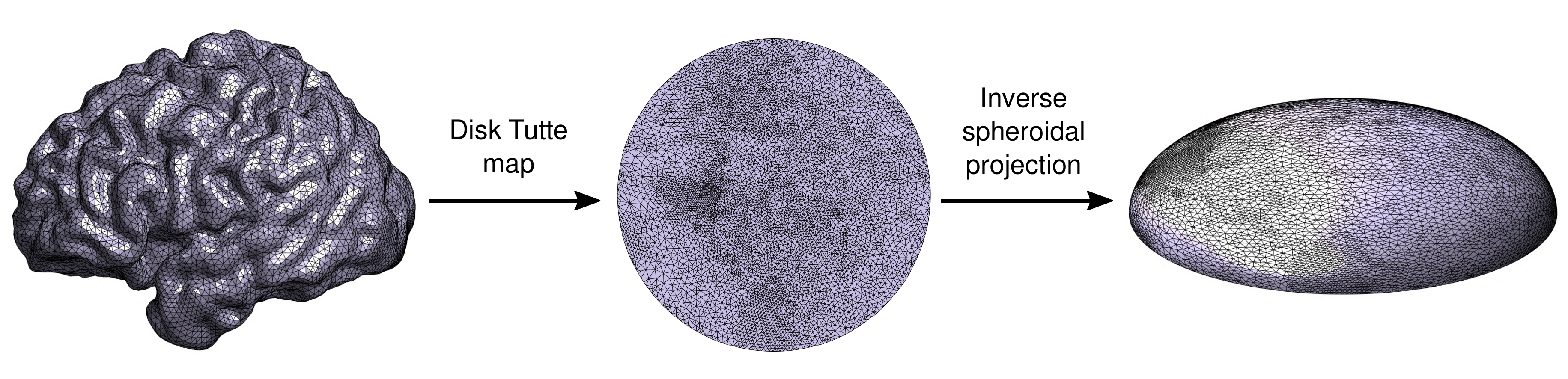}
    \caption{An illustration of our hemispheroidal Tutte parameterization method. We first utilize the Tutte embedding~\cite{tutte1963draw} to construct a disk flattening map. Then, we apply the inverse spheroidal projection to obtain a hemispheroidal parameterization.}
    \label{fig:illustration_tutte}
\end{figure}

\subsection{Hemispheroidal Tutte parameterization}
\label{sect:tutte}
One simple way to compute a bijective map from $\mathcal{M}$ to the target spheroid $\mathcal{H}$ is to utilize the Tutte embedding~\cite{tutte1963draw} (see Fig.~\ref{fig:illustration_tutte} for an illustration of our method). Specifically, we can start by constructing the graph Laplacian $L$, which is a $|\mathcal{V}| \times |\mathcal{V}|$ matrix with
\begin{equation}
    L_{ij} = \left\{\begin{array}{ll}
        1 & \text{ if } [x_i, x_j] \in \mathcal{E}, \\
        \displaystyle -\sum_{\substack{k=1\\k\neq i}}^{|\mathcal{V}|} L_{ik} & \text{ if } i = j, \\
        0 & \text{ otherwise.}
    \end{array}\right.
\end{equation}
Geometrically, the graph Laplacian only focuses on the connectivity of the surface mesh and assumes all edge lengths to be uniform. One can then solve for a bijective map $g:\mathcal{M}\to \mathbb{D}$ from $\mathcal{M}$ to the unit disk $\mathbb{D}$:
\begin{equation} \label{eqt:disk_tutte}
    \left\{\begin{array}{ll}
    L g(v_i) = 0  & \text{ if } v_i \in \mathcal{M} \setminus \partial \mathcal{M} \\
    g(v_{b_j}) = w_{b_j} & \text{ if } v_{b_j} \in  \partial \mathcal{M},
    \end{array}\right.
\end{equation}
where $v_1, v_2, \dots, v_{|\mathcal{V}|}$ are the vertices, $\partial \mathcal{M} = \{v_{b_1},v_{b_1},\dots, v_{b_m}\}$ is the set of all boundary vertices of $\mathcal{M}$, and $w_{b_1},w_{b_2},\dots,w_{b_m}$ are the prescribed target positions of the boundary vertices on the disk boundary. Here, we can set the target positions $w_{b_1},w_{b_2},\dots,w_{b_m}$ based on the edge lengths in the input surface $\mathcal{M}$ as described in~\cite{choi2015fast}. Specifically, we set 
\begin{equation}
    w_{b_j} = e^{i \theta_{b_j}},
\end{equation}
where $l_{[v_{b_t},v_{b_{t+1}}]}$ denotes the length of the boundary edge $[v_{b_t},v_{b_{t+1}}]$ in $\mathcal{M}$, $v_{b_{m+1}} = v_{b_{1}}$, and 
\begin{equation}
    \theta_{b_j} = 2 \pi \left(\sum_{t=1}^j l_{[v_{b_t},v_{b_{t+1}}]}\right) \Big/ \left(\sum_{t=1}^m l_{[v_{b_t},v_{b_{t+1}}]}\right).
\end{equation}
By the above formulation, $w_{b_1},w_{b_2},\dots,w_{b_m}$ will be placed on the unit disk boundary in proportion to the spacing between the corresponding boundary points in the input surface. Because of the use of the graph Laplacian and the convex boundary constraint, the mapping $g$ obtained from Eq.~\eqref{eqt:disk_tutte} is guaranteed to be bijective~\cite{tutte1963draw}.

After obtaining the disk parameterization, we map the result onto the target hemispheroid using a simple projection. It is well-known that the stereographic projection and its inverse are conformal maps between the extended complex plane $\overline{\mathbb{C}}$ and the unit sphere $\mathbb{S}^2$. In particular, they provide a bijection between the unit disk and the Southern hemisphere. One can also consider analogous projection maps between $\overline{\mathbb{C}}$ and an ellipsoid~\cite{choi2024fast}. In our problem, as we want to map the unit disk to the Northern hemispheroid $\mathcal{H}$ in Eq.~\eqref{eqt:hemispheroid}, we define the spheroidal projection $P$ and inverse spheroidal projection $P^{-1}$ as follows:
\begin{equation} \label{eqt:spheroidal_projection}
P(X,Y,Z) = \left(\frac{X}{a\left(1+\frac{Z}{c}\right)}, \frac{Y}{a\left(1+\frac{Z}{c}\right)}\right),
\end{equation} 
\begin{equation} \label{eqt:inverse_spheroidal_projection}
P^{-1}(x,y) = \left(\frac{2ax}{1+x^2+y^2}, \frac{2ay}{1+x^2+y^2}, \frac{-c(-1+x^2+y^2)}{1+x^2+y^2}\right).
\end{equation} 
Then, the final hemispheroidal Tutte parameterization $f_T: \mathcal{M} \to \mathcal{H}$ is given by $f_T = P^{-1} \circ g$.

\subsection{Hemispheroidal conformal parameterization} \label{sect:conformal}
\begin{figure}[t]
    \centering
    \includegraphics[width=0.95\textwidth]{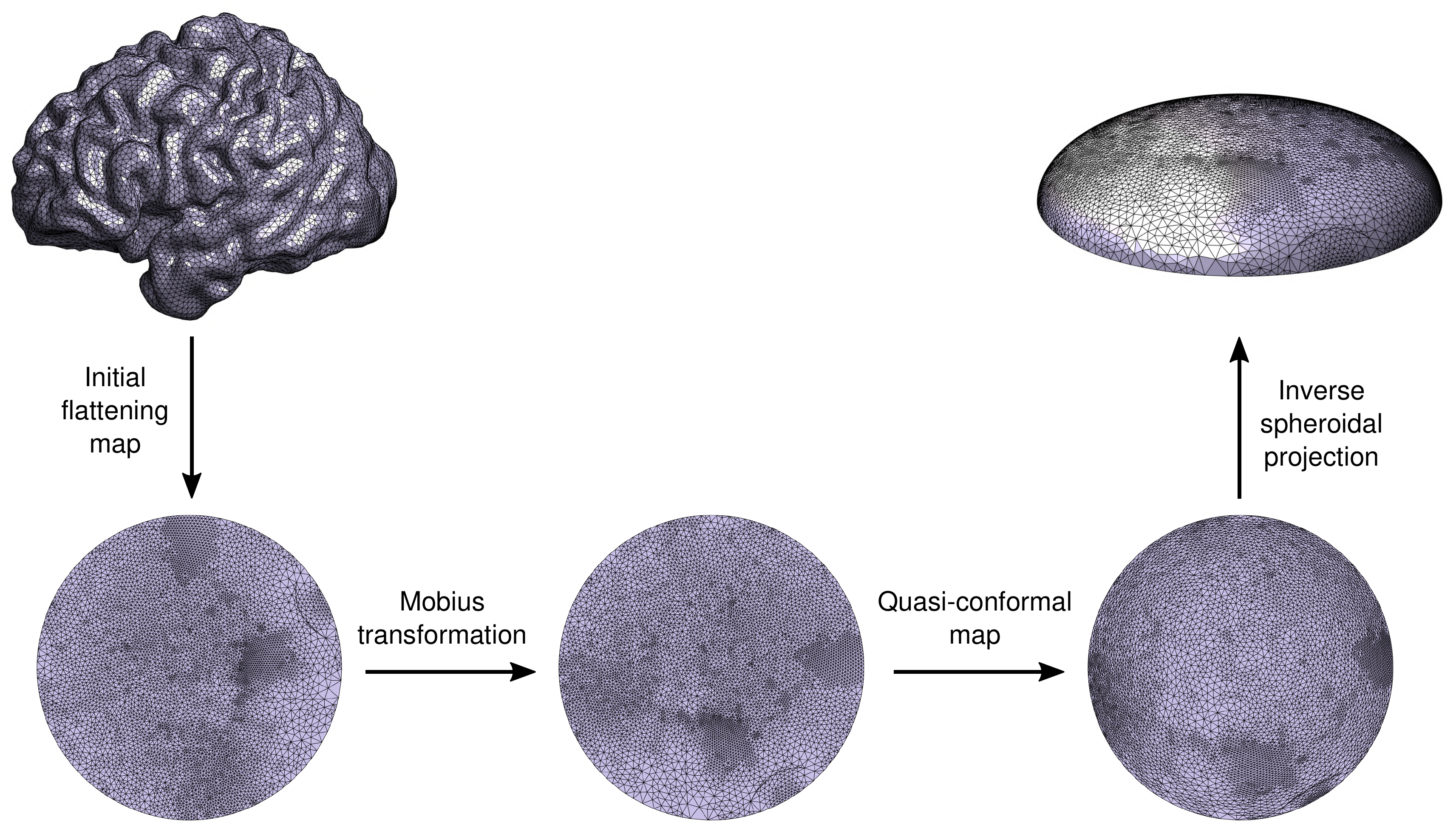}
    \caption{An illustration of our hemispheroidal conformal parameterization method. We start by computing an initial conformal flattening map onto the unit disk followed by an optimal M\"obius transformation. Then, we compose the map with a quasi-conformal map that can effectively correct the conformal distortion caused by the subsequent inverse spheroidal projection. Finally, applying the inverse spheroidal projection gives the desired hemispheroidal conformal parameterization.}
    \label{fig:illustration_conformal}
\end{figure}

While the above-mentioned method can compute a hemispheroidal parameterization efficiently, it does not minimize the geometric distortion in either angle or area. Below, we propose a hemispheroidal conformal parameterization method, which preserves angles and hence the local geometry under the parameterization. Figure~\ref{fig:illustration_conformal} shows an illustration of our proposed method.

\subsubsection{Initial disk conformal parameterization}
We start by flattening the given surface $\mathcal{M}$ onto the unit disk $\mathbb{D}$ using conformal parameterization. Here, we apply the fast disk conformal mapping method~\cite{choi2015fast}, which computes a disk conformal parameterization using an iterative scheme involving a series of analytic functions. Readers are referred to~\cite{choi2015fast} for the details of the method. We denote the mapping by $g_C: \mathcal{M} \to \mathbb{D}$.

It was shown in~\cite{choi2020parallelizable} that one can reduce the area distortion of a conformal parameterization by further composing it with some suitable M\"obius transformations, which are conformal and hence will not affect the conformality of the parameterization. In particular, for disk conformal parameterizations, we can consider automorphisms of the unit disk $\varphi: \mathbb{D} \to \mathbb{D}$ in the form:
\begin{equation}\label{eqt:mobius_proposed}
    \varphi(z) = \frac{z-r e^{i \theta}}{1-r e^{-i \theta}z},
\end{equation}
where $r \in [0,1)$ and $\theta \in [-\pi, \pi)$. Now, since our goal is to obtain a hemispheroidal parameterization, we should focus on reducing the area distortion with the hemispheroidal projection step taken into consideration. To achieve this, we search for an optimal M\"obius transformation $\varphi$ by minimizing the following area distortion energy $E_{\text{area}}$ with respect to the variables $r$ and $\theta$ in Eq.~\eqref{eqt:mobius_proposed}, in which the area of the triangular faces under the inverse spheroidal projection $P^{-1}$ in Eq.~\eqref{eqt:inverse_spheroidal_projection} is considered:
\begin{equation}
    E_{\text{area}} = \frac{1}{|\mathcal{F}|} \sum_{j=1}^{|\mathcal{F}|} \left(\log \frac{\text{Area}((P^{-1} \circ \varphi \circ g_C)(T))}{\sum_{\widetilde{T} \in \mathcal{F}}\text{Area}((P^{-1} \circ \varphi \circ g_C)(\widetilde{T}))}  - \log \frac{\text{Area}(T)}{\sum_{\widetilde{T} \in \mathcal{F}}\text{Area}(\widetilde{T})}\right)^2.
\end{equation}
Note that the denominator in each of the two log terms is used for normalization so that the energy $E_{\text{area}} = 0$ if and only if the mapping $P^{-1} \circ \varphi \circ g_C$ is perfectly area-preserving. In practice, the \texttt{fmincon} function in MATLAB or any other similar optimization solver can be used for solving the above minimization problem. With the optimal $\varphi$ obtained, we can get an updated disk conformal parameterization $\varphi \circ g_C$, with the area distortion of the associated hemispheroidal parameterization result reduced.

\subsubsection{Quasi-conformal composition}
Note that one can obtain a hemispheroidal parameterization by applying the inverse spheroidal projection to the disk conformal parameterization. However, as pointed out in~\cite{choi2024fast}, the generalizations of the stereographic projection and its inverse for spheroids and ellipsoids are not conformal in general. Therefore, the composition map $P^{-1} \circ \varphi \circ g_C$ is not conformal. To get a hemispheroidal conformal parameterization, we need another step of correcting the quasi-conformal distortion of $P^{-1}$. 

Here, we follow the idea in~\cite{choi2024fast} to correct the distortion via quasi-conformal composition. More specifically, we first compute the Beltrami coefficient $\mu_{P^{-1}}$ of the inverse spheroidal projection $P^{-1}$, which is a complex-valued function capturing the local geometric distortion of the mapping (see~\cite{ahlfors2006lectures} for more details). Then, we construct a quasi-conformal map $\psi: \mathbb{D} \to \mathbb{D}$ with the same Beltrami coefficient: $\psi = \textbf{LBS}\left(\mu_{P^{-1}}\right)$, where $\textbf{LBS}(\cdot)$ denotes the quasi-conformal map constructed using the Linear Beltrami Solver (LBS) method~\cite{lui2013texture}, which involves solving a linear system based on a generalized Laplacian matrix. Then, since $\psi$ and $P^{-1}$ have the same Beltrami coefficient, it follows from quasi-conformal theory~\cite{ahlfors2006lectures} that the composition $P^{-1} \circ \psi^{-1}$ is conformal. 

Putting the results together, we obtain a hemispheroidal conformal map $f_C: \mathcal{M} \to \mathcal{H}$ with
\begin{equation}
    f_C = P^{-1} \circ \psi^{-1} \circ \varphi \circ g_C.
\end{equation}
In particular, as the initial disk parameterization $g_C$, the M\"obius transformation $\varphi$, and the composition map $P^{-1} \circ \psi^{-1}$ are all conformal, the final mapping result $f_C$ is also conformal. Also, it is easy to see from their explicit formulas that the M\"obius transformation $\varphi$ and the inverse spheroidal projection $P^{-1}$ are bijective, and the bijectivity of the two other mappings $g_C$ and $\psi$ is guaranteed by their corresponding computational algorithms based on quasi-conformal theory. Therefore, the overall mapping $f_C$ is also bijective.

\begin{figure}[t]
    \centering
    \includegraphics[width=0.95\textwidth]{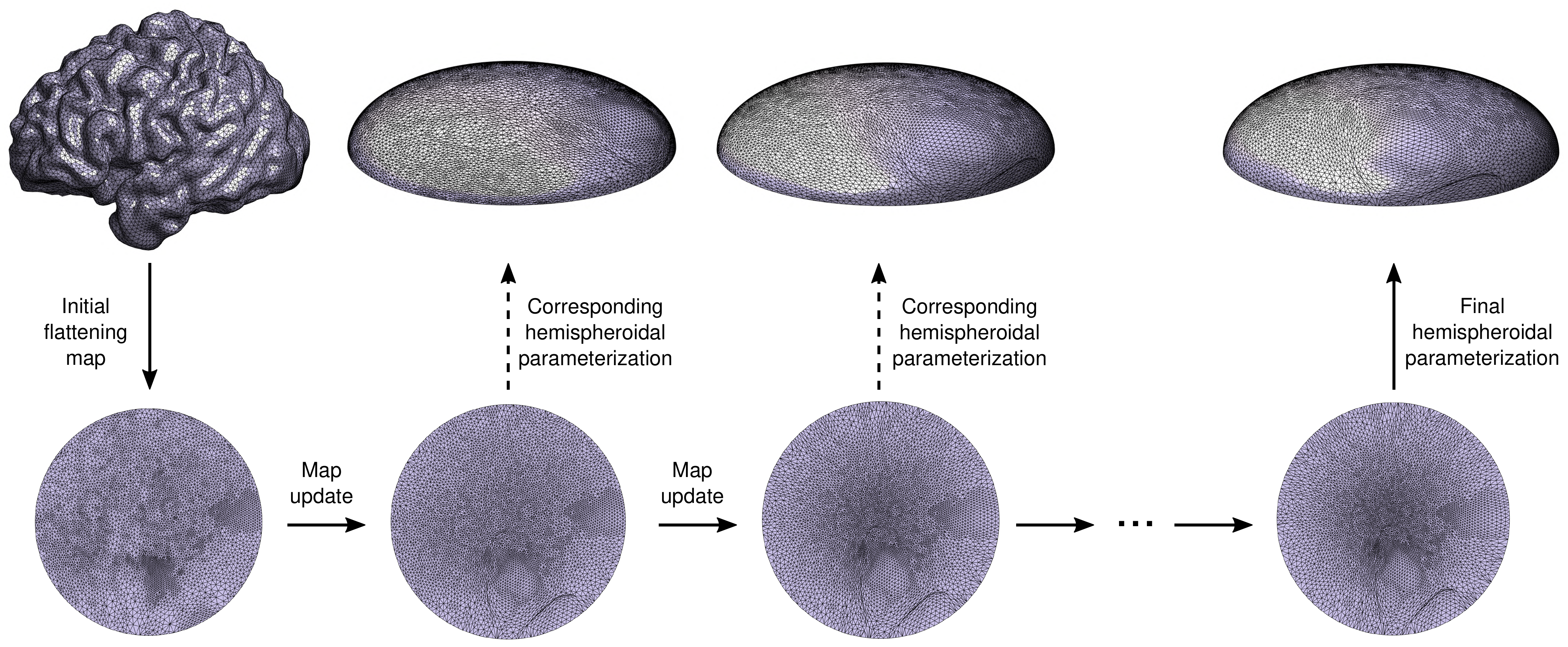}
    \caption{An illustration of our hemispheroidal area-preserving parameterization method. We start by computing an initial surface flattening map onto the unit disk. Then, we perform an iterative update on the planar disk domain based on the disk density-equalizing mapping (DEM) method in~\cite{choi2018density} with certain modifications. Specifically, at each iteration, the mapping update is done on the planar disk domain but the updating criterion is based on the area distortion of the corresponding hemispheroidal map.}
    \label{fig:illustration_AP}
\end{figure}

\subsection{Hemispheroidal area-preserving parameterization}
Besides conformal parameterization, we can also achieve an area-preserving parameterization on the input surface onto the target hemispheroid. To achieve this, we apply the disk density-equalizing mapping (DEM) method in~\cite{choi2018density} with certain modifications. An illustration of our method is given in Fig.~\ref{fig:illustration_AP}.  

The disk DEM method computes a disk area-preserving parameterization by first parameterizing the input simply connected open surface onto the unit disk via a flattening map $g:\mathcal{M} \to \mathbb{D}$. Then, the method deforms the disk domain based on the principle of density diffusion, where the density $\rho$ is set to be 
\begin{equation}
    \rho(T) = \frac{\text{Area}(T)}{\text{Area}(g(T))},
\end{equation}
where $T = [v_i, v_j, v_k]$ is a triangle on the input surface $\mathcal{M}$ and $g(T)$ is the corresponding triangle on the unit disk under the Tutte map $g$. By solving the diffusion equation
\begin{equation} \label{eqt:diffusion}
    \frac{\partial \rho}{\partial t} = \Delta \rho
\end{equation}
and updating the positions of all vertices according to the gradient of $\rho$ iteratively, regions with a larger $\rho$ will be eventually enlarged and regions with a smaller $\rho$ will be shrunk (see~\cite{choi2018density} for the details of the computational procedure). Ultimately, the density is equalized and the resulting map $\tau \circ g$ (where $\tau:\mathbb{D} \to \mathbb{D}$ denotes the density-equalizing map produced by the density equalization process) is a disk area-preserving parameterization. Also, as described in~\cite{shaqfa2023disk}, the local non-bijectivity in the disk parameterization can be corrected by modifying the associated Beltrami coefficient and applying the LBS method~\cite{lui2013texture}.

However, in our hemispheroidal parameterization problem, the ultimate goal is to compute an area-preserving map onto the hemispheroid instead of the unit disk. Therefore, here we first compute a disk Tutte map $g:\mathcal{M} \to \mathbb{D}$ using Eq.~\eqref{eqt:disk_tutte}. We then further apply a M\"obius transformation $\varphi$ in Eq.~\eqref{eqt:mobius_proposed} to reduce the area distortion as described in Section~\ref{sect:conformal}. Next, for the density-equalizing mapping step, instead of the density $\rho$ above, we set the density $\rho_{H}$ based on the triangle area \emph{under the inverse spheroidal projection $P^{-1}$} as follows:
\begin{equation} \label{eqt:density_H}
    \rho_{H}(T) = \frac{\text{Area}(T)}{\text{Area}(P^{-1}((\varphi \circ g)(T)))}.
\end{equation}
Then, although the density diffusion process will take place on the unit disk, the composition of the updated disk mapping result and the inverse spheroidal projection will give a hemispheroidal area-preserving map. More specifically, denote the density-equalizing map obtained by solving the diffusion equation~\eqref{eqt:diffusion} based on $\rho_H$ as $\tau_{H}:\mathbb{D} \to \mathbb{D}$. The final hemispheroidal area-preserving parameterization $f_A: \mathcal{M} \to \mathcal{H}$ is given by 
\begin{equation}
    f_A = P^{-1} \circ \tau_{H}\circ \varphi \circ g.
\end{equation}
We remark that analogous to the disk parameterization problem~\cite{shaqfa2023disk}, the local mesh fold-overs produced throughout the density diffusion process in our hemispheroidal parameterization problem can also be corrected using the LBS method.
 
Besides, as mentioned in a recent work~\cite{lyu2024spherical}, numerical errors may accumulate throughout the density equalization process due to small discrepancies between the density and the mapping result. Therefore, here we also apply the re-coupling scheme proposed in~\cite{lyu2024spherical} at the end of each iteration throughout the density diffusion process. More specifically, instead of directly solving Eq.~\eqref{eqt:diffusion} based on $\rho_{H}$ with the fixed denominator $\text{Area}(P^{-1}((\varphi \circ g)(T)))$ in Eq.~\eqref{eqt:density_H}, we re-define $\rho_{H}$ as 
\begin{equation}
   \rho_H(T) =  \frac{\text{Area}(T)}{\text{Area}(P^{-1}([\mathbf{x}_i,\mathbf{x}_j,\mathbf{x}_k]))},
\end{equation}
where $ \mathbf{x}_i,\mathbf{x}_j,\mathbf{x}_k$ are the latest updated vertex positions corresponding to the vertices $v_i, v_j, v_k$ at the end of each iteration. The update of the vertex positions in the next iteration is then based on the gradient of the re-defined $\rho_H$. This additional step ensures that numerical errors will not accumulate throughout the iterative process, thereby improving the accuracy of the hemispheroidal area-preserving parameterization.

\subsection{Hemispheroidal parameterization for balancing different distortions}
In some situations, it may be desirable for the user to customize the hemispheroidal parameterization with a balance between different distortions. To achieve this, we can utilize the correspondence between Beltrami coefficients and mappings in quasi-conformal theory. More specifically, we start by computing the disk Tutte map $g:\mathcal{M} \to \mathbb{D}$ as described in Section~\ref{sect:tutte}. Then, we compute the Beltrami coefficients of the mappings from $\mathbb{D}$ to the parameterization results obtained by the three above-mentioned methods right before the final inverse spheroidal projection (i.e. $P \circ f_T$, $P \circ f_C$, and $P \circ f_A$) and denote them as $\mu_T$, $\mu_C$, and $\mu_A$, respectively. Now, we consider the following balanced Beltrami coefficient $\mu_B$: 
\begin{equation}
    \mu_B = \alpha \mu_T + \beta \mu_C + \gamma \mu_A,
\end{equation}
where $\alpha$, $\beta$, $\gamma$ are three nonnegative weighting parameters with $\alpha + \beta + \gamma = 1$. Then, we can utilize the LBS method~\cite{lui2013texture} to reconstruct a quasi-conformal map $\psi_B: \mathbb{D} \to \mathbb{D}$ with the same Beltrami coefficient:  $\psi_B = \textbf{LBS}\left(\mu_B\right)$. Finally, we apply the inverse spheroidal projection to obtain a hemispheroidal parameterization $f_B: \mathcal{M} \to \mathcal{H}$:
\begin{equation}
    f_B = P^{-1} \circ \psi_B \circ g.
\end{equation}
It is easy to see that if $\alpha$ is much larger than $\beta$ and $\gamma$, the combined Beltrami coefficient $\mu_B$ will be close to $\mu_T$ and hence the final hemispheroidal parameterization result $f_B$ will be close to the hemispheroidal Tutte parameterization $f_T$. Similarly, if $\beta$ is much larger then $f_B$ will be close to the hemispheroidal conformal parameterization $f_C$, and if $\gamma$ is much larger then $f_B$ will be close to the hemispheroidal area-preserving parameterization $f_A$. A balance between $\alpha$, $\beta$, and $\gamma$ will then yield a hemispheroidal parameterization $f_B$ balancing different distortions.

\section{Harmonics functions} \label{sect:harmonics}
\subsection{Spheroidal coordinates}
\label{subsec:coords}
\par
As described in Shaqfa and van Rees~\cite{shaqfa2024spheroid}, spheroids can be classified into oblates and prolates based on the aspect ratio $\text{AR} = a/c$, where $a$ is the repeated semi-axis, meaning the intermediate semi-axis $b \coloneq a$, and $c$ is the remaining one. For $\text{AR} > 1$ we have an oblate hemispheroid, while for the prolate one $\text{AR}\leq 1$.
\par
\begin{figure}[t]
    \centering
    \includegraphics[width=0.75\textwidth]{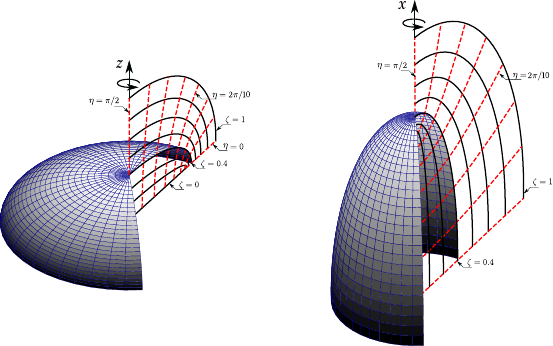}
    \caption{Hemispheroidal spaces generated as a revolution of elliptic coordinates. Revolving about the z-axis will result in an oblate space (left inset), while revolving about the x-axis will result in an oblate space (right inset).}
    \label{fig:hemispheroid_coord}
\end{figure}

From \cite{byerly1959elementary, Gray1997, Moon1988}, we can summarize the parametric form of the confocal oblate coordinates as:
\begin{equation}\label{eqn:oblate_para1}
    \begin{aligned}
        x(\zeta, \eta, \phi) & = e \ \cosh \zeta \ \cos \eta \ \cos \phi,
        \\ y(\zeta, \eta, \phi) & = e \ \cosh \zeta \ \cos \eta \ \sin \phi,
        \\ z(\zeta, \eta) & = e \ \sinh \zeta \ \sin \eta.
    \end{aligned}
\end{equation}
Here, $e$ is the focal distance measured from the origin, $\zeta$ is the confocal elliptic section, and $\eta$ is a hyperbolic section that is analogous to the radial distance of spherical coordinates and it is orthogonal to $\zeta$. For the basis functions $\eta$ is the latitude angle and its domain is $[-\pi/2, \pi/2]$. The azimuthal angle $\phi$ is defined over the domain $[0, 2 \pi]$. An alternative notation for Eq.~\eqref{eqn:oblate_para1} is by assuming $\xi_1 = \sinh \zeta$ and $\xi_2 = \sin \eta$. The latter notation, as a function of $\xi_1$ and $\xi_2$, is important to write the general Laplacian equation in the spheroidal coordinates. For the prolate coordinates, we write the parametric form as:
\begin{equation}\label{eqn:prolate_para1}
    \begin{aligned}
        x(\zeta, \eta, \phi) & = e \ \sinh \zeta \ \sin \eta \ \cos \phi,
        \\ y(\zeta, \eta, \phi) & = e \ \sinh \zeta \ \sin \eta \ \sin \phi,
        \\ z(\zeta, \eta) & = e \ \cosh \zeta \ \cos \eta.
    \end{aligned}
\end{equation}
For the prolate case, the latitude angle $\eta$ range is $[0, \pi]$. And the alternative form for the coordinates is written by assuming $\xi_1 = \cosh \zeta$ and $\xi_2 = \cos \eta$. Figure \ref{fig:hemispheroid_coord} shows the coordinates of revolution of the hemispheroidal space for oblate and prolate cases.
\par
For hemispheroidal coordinates, we change the domain of $\eta$ depending on whether we choose to use the Northern or Southern hemispheroid. We here use the Northern hemispheroid in our basis functions, where the latitude angle $\eta \in [0, \pi/2]$ for both oblate and prolate Northern hemispheroids. The computation of the inverse coordinates $(\zeta, \eta, \phi)$ from the Cartesian ones $(x, y, z)$ will not be discussed here and we refer readers to \cite{shaqfa2024spheroid} as we proposed in that paper the usage of two different inversions that will give the same results for computing $\eta$ and $\phi$ when the points are laid onto a spheroid/hemispheroid. Due to the overflow in the basis functions, having $\eta \to \pi/2$ could cause a faulty reconstruction near the edge of open surfaces, so we here assume $\max(\eta) = \pi/2 - \epsilon_{\eta}$ such that $\epsilon_{\eta}$ is a real nonzero positive number. The $\epsilon_{\eta}$ value has changed in each one of the reconstructed examples (see the subsequent section) such that we do not truncate part of the near-edge surface nor have severely oscillating edges.

\subsection{Hemispheroidal harmonic (HSOH) basis functions}
\par
In Shaqfa et~al. (2021)~\cite{shaqfa2021spherical}, we used the spherical cap harmonics basis to decompose open surfaces. Although that approach was robust and allowed for a flexible parameterization, it suffered from numerical instabilities with the basis that hindered the analysis from achieving high expansion degrees. In this work, we propose the usage of hemispheroidal harmonics to expand single-edged parametric surfaces with the same degree of customizability enabled in the SCHA \cite{shaqfa2021spherical}, however, using hemispheroidal analysis domain.
\par
The general Laplacian operator in the confocal spheroidal coordinates, by using the $\xi_1$ and $\xi_2$ notations, can be written as:
\begin{equation}
\label{subsec:hemi_basis_funcs}
    \nabla^{2} f = \frac{1}{e^{2}(\xi_1^{2} \pm \xi_2^{2})}\left(\frac{\partial}{\partial\xi_1}\left((\xi_1^{2} \pm 1)\frac{\partial f}{\partial\xi_1}\right)+\frac{\partial}{\partial\xi_2}\left((1-\xi_2^{2})\frac{\partial f}{\partial\xi_2}\right)+\frac{\xi_1^{2} \pm \xi_2^{2}}{(\xi_1^{2} \pm 1)(1-\xi_2^{2})}\frac{{\partial^{2} f}}{{\partial\phi}^{2}}\right).
\end{equation}
Notice the $\pm$ signs associated with the $\xi_1$ terms account for the oblate and prolate cases. The upper sign of $\pm$ is for the Laplacian operator of the oblate spheroid and the lower one is for the prolate spheroid. Similar to the traditional spherical harmonics, the separable solution of this elliptic Laplacian operator can be found in classical literature as a multiplication of the associated Legendre polynomials and Fourier basis.
\par
The spheroidal harmonic bases were used for decomposing closed genus-0 parametric surfaces in Shaqfa and van Rees (2024) \cite{shaqfa2024spheroid}. Similarly, we here use the same basis functions with a slight modification to account for the edge of the hemispheroid with an applied Neumann boundary condition on it. The used basis functions of a spheroidal surface, given that $\zeta = const.$, can be expressed as:
\begin{equation}\label{eqn:general_sol}
    f^n_m (\eta, \phi) = N_m^n \ P_m^n\left(\hat{\xi}\right) \ e^{i m \phi}.
\end{equation}
\par
\begin{figure}[!t]
    \centering
    \includegraphics[width=0.75\linewidth]{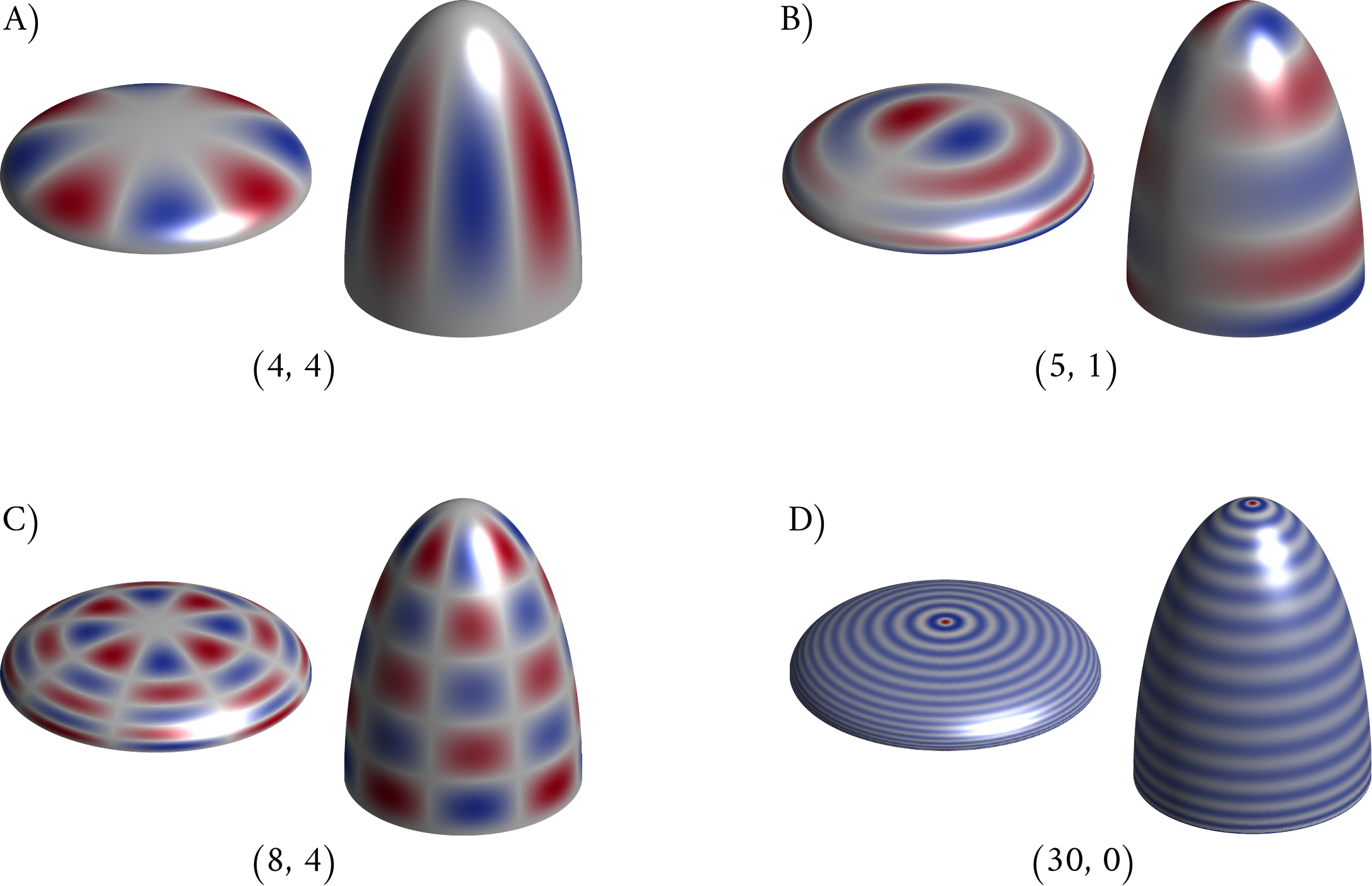}
    \caption{Hemishpheroidal harmonics basis functions for a selected set of $(n, m)$. The shown heat maps represent the normalized real part of the basis $\ \Re\{ N_m^n \ S^n_m(\eta, \phi) \}$ for $\zeta = 0.4$ and $ e = 1$.}
    \label{fig:hemispheroidal_harm_basis}
\end{figure}

Analogous to the hemispherical harmonics, we here introduce an affine transformation of the associated Legendre polynomials. This transformation allows for the shifting of the orthogonality domain depending on the size of the hemispheroid. From that, it follows that $\hat{\xi}$ is written as:
\begin{equation}
    \hat{\xi} = 
    \left\{\begin{matrix}
    ~~2 \sin \eta - 1 & \qquad \text{AR} > 1, \\ 
    -2 \cos \eta + 1 & \qquad \text{AR} \leq 1.
    \end{matrix}\right.
\end{equation}
Here, $n$ and $m$ are the degree and order of the solutions, with $\hat{\xi} \in [-1, 1]$ satisfying the orthogonality domain for the ALP. $\eta$ and $\phi$ are the latitude and longitude coordinates of the spheroid, respectively. Let $\mathbf{f} (\eta, \phi) = [{x}(\eta, \phi), {y}(\eta, \phi), {z}(\eta, \phi)]^{T}$ represent the surface data in the parametric form of a spheroidal space $\mathcal{H_S}$. $N_m^n$ is a normalization factor to avoid the data overflow of high-degree harmonics and make the basis orthonormal. Then, the harmonic decomposition of a spheroidal surface can be written as:
\begin{equation}\label{eqn:general_decomp}
    \mathbf{f}(\eta, \phi) = \sum_{n=0}^{\infty} \sum_{m=-n}^{n} A_m^n \ N_m^n \ P_m^n\left(\hat{\xi}\right) \ e^{i m \phi} = \sum_{n=0}^{\infty} \sum_{m=-n}^{n} A^n_m S^n_m.
\end{equation}
$A_m^n$ are the linear Fourier weights of the expansion that are computed as $\langle \mathbf{f}(\eta, \phi), S^n_m \rangle$, where $\langle \cdot, \cdot \rangle$ is the inner product. Many ways can be used to efficiently compute these weights (see \cite{Shaqfa2023OnMethod} for details). Figure \ref{fig:hemispheroidal_harm_basis} shows the real part of the surface basis functions of the hemispheroidal harmonics.
Practically, a parametric surface can be approximated by a maximum number of expansion degrees $n_{max}$ such that $\hat{\mathbf{f}}(\eta, \phi) \approx \mathbf{f}(\eta, \phi)$, where:
\begin{equation}\label{eqn:general_decomp_final}
    \hat{\mathbf{f}}(\eta, \phi) = \sum_{n=0}^{n_{max}} \sum_{m=-n}^{n} A^n_m S^n_m.
\end{equation}

\section{Experiments} \label{sect:experiments}
In this section, we summarize the results of the herein-proposed parameterization approaches and the harmonic decomposition. We start with a detailed analysis of our proposed hemispheroidal parameterization methods in terms of the geometric distortion of the parameterization results, followed by a comparison between the proposed approaches and prior disk and hemispherical parameterization methods. We then examine the effect of our proposed hemispheroidal parameterization and harmonic decomposition framework on surface reconstruction. We also briefly compare the results with our recent papers, namely, the disk harmonics (DH) and the spherical cap harmonics (SCH). Lastly, we show the reconstruction effects when optimizing the size of the hemispheroid, as well as, the optimum Beltrami-based parameterization results.

\subsection{Hemispheroidal parameterization}
\label{subsec:parameterization_results}
\par

\begin{figure}[t]
    \centering
    \includegraphics[width=\textwidth]{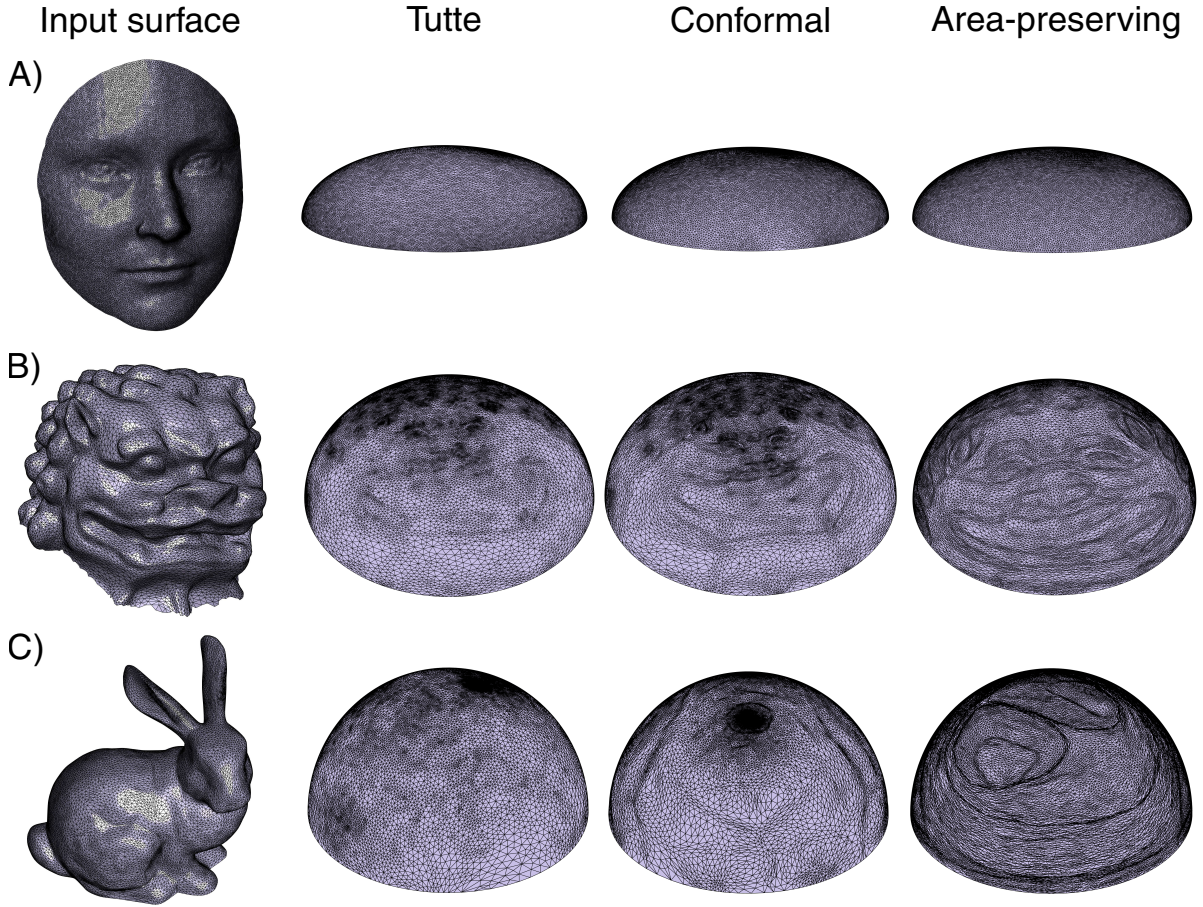}
    \caption{Examples of simply connected open surfaces and the hemispheroidal parameterization results. (A) The human face model. (B) The Chinese Lion model. (C) The Stanford bunny model. For each example, we show the input surface, the hemispheroidal Tutte parameterization, the hemispheroidal conformal parameterization, and the hemispheroidal area-preserving parameterization.}
    \label{fig:mapping_result}
\end{figure}

In Fig.~\ref{fig:mapping_result}, we show several examples of simply connected open surfaces and the results of the hemispheroidal Tutte parameterization, hemispheroidal conformal parameterization, and hemispheroidal area-preserving parameterization obtained by our proposed approaches. As shown, the three hemispheroidal parameterization methods exhibit different outcomes. To quantitatively assess the geometric distortion of the parameterizations, we evaluate the angle and area distortion metrics. For any given parameterization $f$, the angle distortion at every angle $\angle [v_i, v_j, v_k]$ (in degree) is evaluated as:
\begin{equation} \label{eqt:angle}
    d_{\text{angle}}(\angle [v_i, v_j, v_k]) = \angle [f(v_i), f(v_j), f(v_k)] -  \angle [v_i, v_j, v_k].
\end{equation}
For a perfectly conformal parameterization $f$, $d_{\text{angle}} = 0$ everywhere. As for the area distortion, for any triangle $T = [v_i, v_j, v_k]$, we evaluate the area distortion of $f$ as follows:
\begin{equation} \label{eqt:area}
    d_{\text{area}}(T) = \log\left(\frac{\text{Area}([f(v_i),f(v_j),f(v_k)])/A_f}{\text{Area}([v_i,v_j,v_k])/A_o}\right),
\end{equation}
where $A_o = \sum_{\tilde{T}} \text{Area}(\tilde{T})$ and $A_f = \sum_{\tilde{T}} \text{Area}(f(\tilde{T}))$ are the triangle area sum of the input surface $\mathcal{M}$ and the parameterization result $f(\mathcal{M})$, which are used as normalization factors here. For a perfectly area-preserving map $f$, $d_{\text{area}} = 0$ for all triangular faces.

\par
Table~\ref{tab:comparison_hemispheroidal} compares different parameterization results across multiple open surfaces. From the same table, we notice that the hemispheroidal conformal parameterization achieves minimal angle distortion but relatively large area distortion, and vice versa for the hemispheroidal area-preserving parameterization. While the computational procedure of the hemispheroidal Tutte parameterization is the simplest and only involves solving a linear system, it is neither conformal nor area-preserving. 

\begin{table}[t]
    \centering
    
\resizebox{0.95\textwidth}{!}{%
    \begin{tabular}{|c|c|c|c|c|c|c|} \hline
       \multirow{2}{*}{\textbf{Surface}}  & \multicolumn{2}{c|}{\textbf{Tutte}} & \multicolumn{2}{c|}{\textbf{Conformal}} & \multicolumn{2}{c|}{\textbf{Area-Preserving}}\\\cline{2-7}
       & $\left|d_{\text{angle}}\right|$ & $\left|d_{\text{area}}\right|$ & $\left|d_{\text{angle}}\right|$ & $\left|d_{\text{area}}\right|$ & $\left|d_{\text{angle}}\right|$ & $\left|d_{\text{area}}\right|$  \\ \hline
       Face & 20.55$\pm$15.66 & 0.67$\pm$0.42 & {\bf 0.49$\pm$0.63} & 0.34$\pm$0.38 &   10.69$\pm$8.98  & {\bf 0.04$\pm$0.07}\\ \hline
       Sophie & 18.03$\pm$13.57 & 0.52$\pm$0.34  & {\bf 0.34$\pm$0.59} & 0.22$\pm$0.30 & 10.48$\pm$8.72 & {\bf 0.07$\pm$0.06} \\ \hline
       Julius & 20.64$\pm$15.26 & 0.64$\pm$0.42 & {\bf 0.85$\pm$2.21} & 0.52$\pm$0.45 & 14.44$\pm$11.60 & {\bf 0.03$\pm$0.05} \\  \hline
       Chinese Lion & 13.02$\pm$9.99  & 0.68$\pm$0.48 & {\bf 1.39$\pm$2.04} & 0.75$\pm$0.48 & 20.05$\pm$15.03 & {\bf 0.11$\pm$0.09}\\ \hline
       Matterhorn& 27.92$\pm$21.00 & 0.69$\pm$0.49 & {\bf 0.38$\pm$0.80} & 0.43$\pm$0.59 & 19.45$\pm$15.16 & {\bf 0.05$\pm$0.04} \\ \hline
       Brain & 17.77$\pm$13.70 & 0.50$\pm$0.36 & {\bf 1.86$\pm$2.68} & 0.36$\pm$0.30  & 12.82$\pm$10.77 & {\bf 0.04$\pm$0.07} \\ \hline
       Bunny & 19.67$\pm$16.54 & 2.35$\pm$2.62 & {\bf 1.22$\pm$1.51} & 2.84$\pm$4.15 & 35.18$\pm$27.16 & {\bf 0.15$\pm$0.24} \\ \hline
    \end{tabular}
    }
    \caption{Comparison between our proposed hemispheroidal Tutte, conformal, and area-preserving parameterizations. For each method, we evaluate the absolute angle distortion $\left|d_{\text{angle}}\right|$ [where $d_{\text{angle}}$ is the angle difference in degree as in Eq.~\eqref{eqt:angle}] and absolute area distortion $\left|d_{\text{area}}\right|$ [where $d_{\text{area}}$ is the logged area ratio as in Eq.~\eqref{eqt:area}]. Each column shows the mean $\pm$ standard deviation of the distortion, and the best result for each type of distortion for each example is boldfaced.}
    \label{tab:comparison_hemispheroidal}
\end{table}

It is natural to ask whether the use of the hemispheroidal domain for the parameterizations is advantageous over other commonly used domains such as the unit disk or the unit hemisphere. Here, we compare the hemispheroidal conformal parameterization with the disk conformal parameterization~\cite{choi2015fast} and the hemispherical conformal map (obtained using the disk conformal map~\cite{choi2015fast} followed by the inverse stereographic projection. For the hemispheroidal parameterization results, we assumed the radius $c$ to be determined by the registration process in Section \ref{subsec:surface_registraction}. As shown in Table~\ref{tab:comparison_conformal}, the proposed hemispheroidal conformal parameterization method achieves similar and sometimes better angle-preserving properties when compared to the two other methods. This can be explained by the fact that the latter methods use a fixed disk or hemisphere as the parameter domain and may only be suitable for certain nominally flat surfaces (such as human faces or rough self-affine fractal surfaces \cite{shaqfa2023disk}) or surfaces close to a hemispherical shape (such as the Chinese lion surface) respectively. By contrast, the flexibility of our hemispheroidal domain allows us to handle a wide range of surfaces more easily. Analogously, we can compare our proposed hemispheroidal area-preserving parameterization with the disk area-preserving parameterization method~\cite{zhao2013area} and the hemispherical area-preserving parameterization method~\cite{giri2021open}. From the results in Table~\ref{tab:comparison_area}, we see that our hemispheroidal method effectively reduces the area distortion when compared to the two other approaches. The further improved mapping qualities suggest the significance of using the hemispheroidal domain with an additional degree of freedom $c$ for the parameterizations.

\begin{table}[t]
    \centering
\resizebox{0.7\textwidth}{!}{%
    \begin{tabular}{|c|c|c|c|} \hline
       \textbf{Surface}  & {\textbf{Disk}}~\cite{choi2015fast} & \textbf{Hemispherical}~\cite{choi2015fast} & \textbf{Hemispheroidal}\\ \hline
       Face & {\bf 0.49$\pm$0.63} &  0.53$\pm$0.63 &  {\bf 0.49$\pm$0.63} \\ \hline
       Sophie & 0.34$\pm$0.60  & 0.39$\pm$0.58 &  {\bf 0.34$\pm$0.59} \\ \hline
       Julius & {\bf 0.85$\pm$2.19} & 0.87$\pm$2.21  & 0.85$\pm$2.21 \\ \hline
       Chinese Lion & 1.42$\pm$2.04 & {\bf 1.37$\pm$2.04} & 1.39$\pm$2.04\\ \hline
       Matterhorn & {\bf 0.38$\pm$0.80} & 0.39$\pm$0.80 & {\bf 0.38$\pm$0.80} \\ \hline
       Brain & 1.87$\pm$2.66 & {\bf 1.86$\pm$2.68} & {\bf 1.86$\pm$2.68}\\ \hline
       Bunny & 1.32$\pm$1.51 & {\bf 1.22$\pm$1.51} & {\bf 1.22$\pm$1.51} \\ \hline
    \end{tabular}
    }
    \caption{Comparison between disk conformal parameterization~\cite{choi2015fast}, hemispherical conformal parameterization (using~\cite{choi2015fast} with an inverse stereographic projection) and our proposed hemispheroidal conformal parameterization method. For each method, we evaluate the absolute angle distortion $\left|d_{\text{angle}}\right|$ to assess its angle-preserving property. Each column shows the mean $\pm$ standard deviation of the distortion. The best result for each example is boldfaced.}
    \label{tab:comparison_conformal}
\end{table}

\begin{table}[t]
    \centering
\resizebox{0.7\textwidth}{!}{%
    \begin{tabular}{|c|c|c|c|} \hline
       \textbf{Surface}  & \textbf{Disk}~\cite{zhao2013area} & \textbf{Hemispherical}~\cite{giri2021open} & \textbf{Hemispheroidal}\\ \hline
       Face & 0.18$\pm$0.22 &  0.25$\pm$0.26 & {\bf 0.04$\pm$0.07} \\ \hline
       Sophie & 0.11$\pm$0.09 & 0.11$\pm$0.09 & {\bf 0.07$\pm$0.06} \\ \hline
       Julius & 0.26$\pm$0.24 & 0.31$\pm$0.27  & {\bf 0.03$\pm$0.05}  \\ \hline
       Chinese Lion &  0.17$\pm$0.20 & 0.16$\pm$0.15 & {\bf 0.11$\pm$0.09} \\ \hline
       Matterhorn& 0.42$\pm$0.54 & 0.54$\pm$0.68 & {\bf 0.05$\pm$0.04}  \\ \hline
       Brain & 0.07$\pm$0.10  &  0.07$\pm$0.10  & {\bf 0.04$\pm$0.07} \\ \hline
       Bunny & 0.21$\pm$0.35 & 0.18$\pm$0.25 & {\bf 0.15$\pm$0.24} \\ \hline
    \end{tabular}
    }
    \caption{Comparison between disk area-preserving parameterization~\cite{zhao2013area}, hemispherical area-preserving parameterization~\cite{giri2021open} and our proposed hemispheroidal area-preserving parameterization method. For each method, we evaluate the absolute absolute area distortion $\left|d_{\text{area}}\right|$ to assess its area-preserving property. Each column shows the mean $\pm$ standard deviation of the distortion. The best result for each example is boldfaced.}
    \label{tab:comparison_area}
\end{table}

\subsection{Hemispheroidal harmonics (HSOH) analysis and reconstruction}
\label{subsec:reconstruction_results}
\par
As we demonstrated the improvement in the angle and area distortions achieved by our proposed parameterizations, we now pair our proposed methods with the hemispheroidal harmonic (HSOH) decomposition and compare their reconstruction results with existing methods. The success of the parametric surface decomposition relies on the existence of a bijective image mapped onto the parameterization domain. Prior works considered traditional mapping methods and combined them with the harmonic basis functions defined on the associated domain for the reconstruction. While, how the mapping choice affects the reconstruction result and how one can determine an optimal mapping for the reconstruction are less understood. More specifically, conventional parameterization methods are not tailored for preserving the orthogonality of the sampled basis onto the target decomposition domain. Indeed, preserving the orthogonality of the basis requires a careful layout of the mapped points onto a mathematical lattice \cite{Shaqfa2023OnMethod}. In the following, we consider the proposed parameterization approaches and carefully compose them to maximize the reconstruction accuracy.
\par
The herein-studied benchmarks are taken from the literature of simply connected open surfaces. We analyzed and reconstructed the surfaces via the HSOH basis functions paired with one or more of the proposed parameterization approaches. Based on the parameterization method choice, we denote the Tutte mapping-based decomposition by T-HSOH, the conformal mapping-based one by C-HSOH, the area-preserving mapping-based one by AP-HSOH, and finally the balanced mapping-based decomposition by B-HSOH. To measure the reconstruction error, we used the mean Hausdorff distance between the input and output meshes and denoted this error measure by A-RMSE.

\subsubsection{Hierarchical harmonic reconstruction of complex geometries}
\par
One important application for surface harmonic decomposition is the frequency-based morphology of a given surface. Here, we demonstrated this ability of our proposed hemispheroidal parameterization and harmonics framework to an anatomical open surface of a brain hemisphere. Successful decomposition of anatomical surfaces depends on preserving the surface's local morphological features; thus, area-preserving mapping is a prime candidate. 
\par
\begin{figure}[!t]
    \centering
    \includegraphics[width=\linewidth]{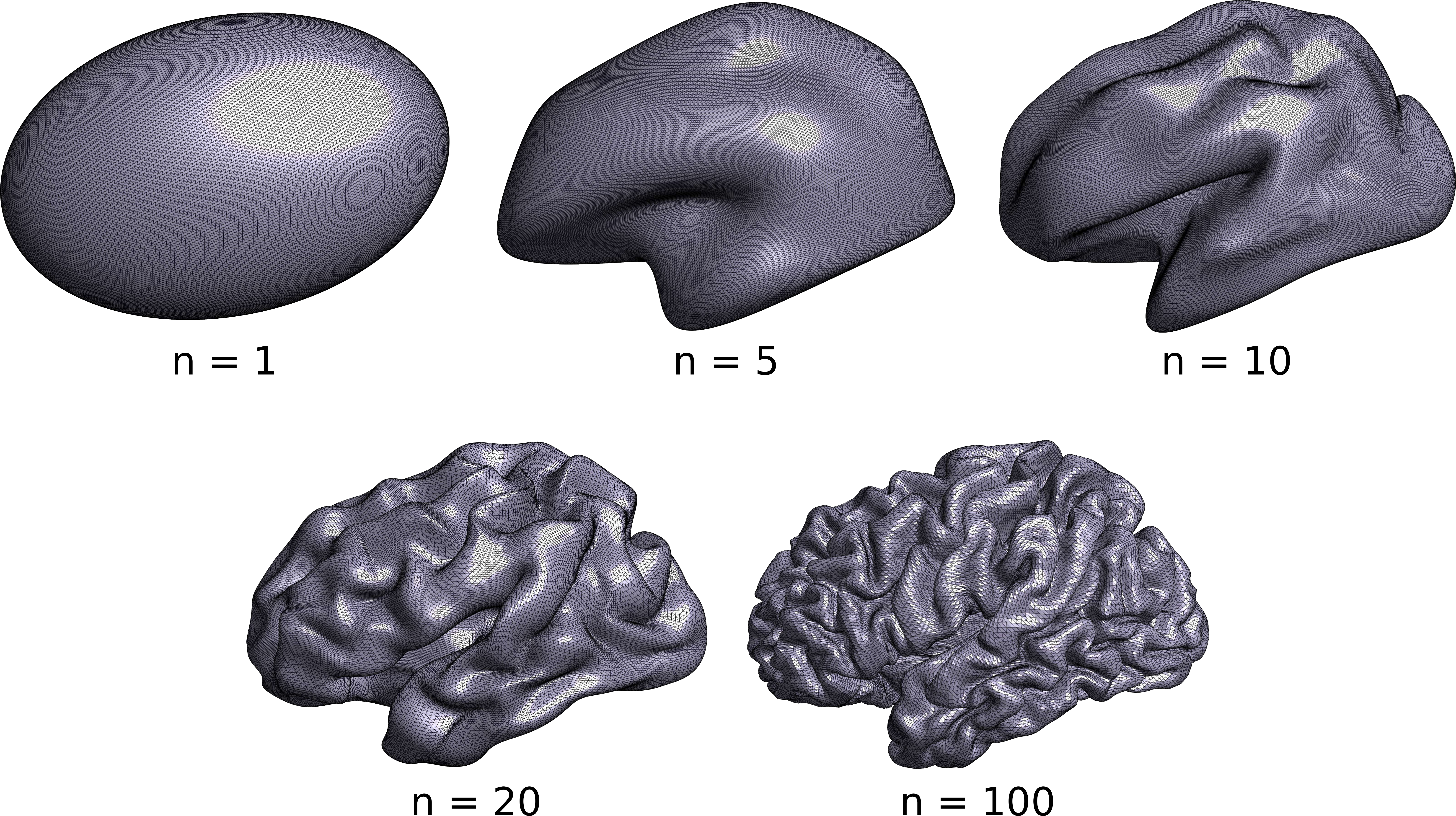}
    \caption{Reconstruction of the human half brain cortical surface at different degrees ($n$) with $n_{max} = 100$, where the input surface is constituted of $48,463$ vertices. The parameterization domain constitutes an oblate hemispheroid with a size of $a = 1, \text{ and } c = 0.5343$. Using the AP-HSOH approach, we obtained the reconstruction error A-RMSE = $0.001318$. The surface was reconstructed using a hemispheroid (of the same size) with $36,268$ uniformly distributed vertices.}
    \label{fig:brain_rec}
\end{figure}

Using the AP-HSOH approach with $n_{max} = 100$, we achieved a reconstructed accuracy of A-RMSE = $0.001318$. Figure~\ref{fig:brain_rec} shows the hierarchical harmonic reconstruction of the cortical surface, in which we can see that our hemispheroidal framework can effectively reconstruct a complex surface with different levels of detail.

\subsubsection{Preservation of sharp features in the harmonic reconstruction}
\par
\begin{figure}[!t]
    \centering
    \includegraphics[width=\linewidth]{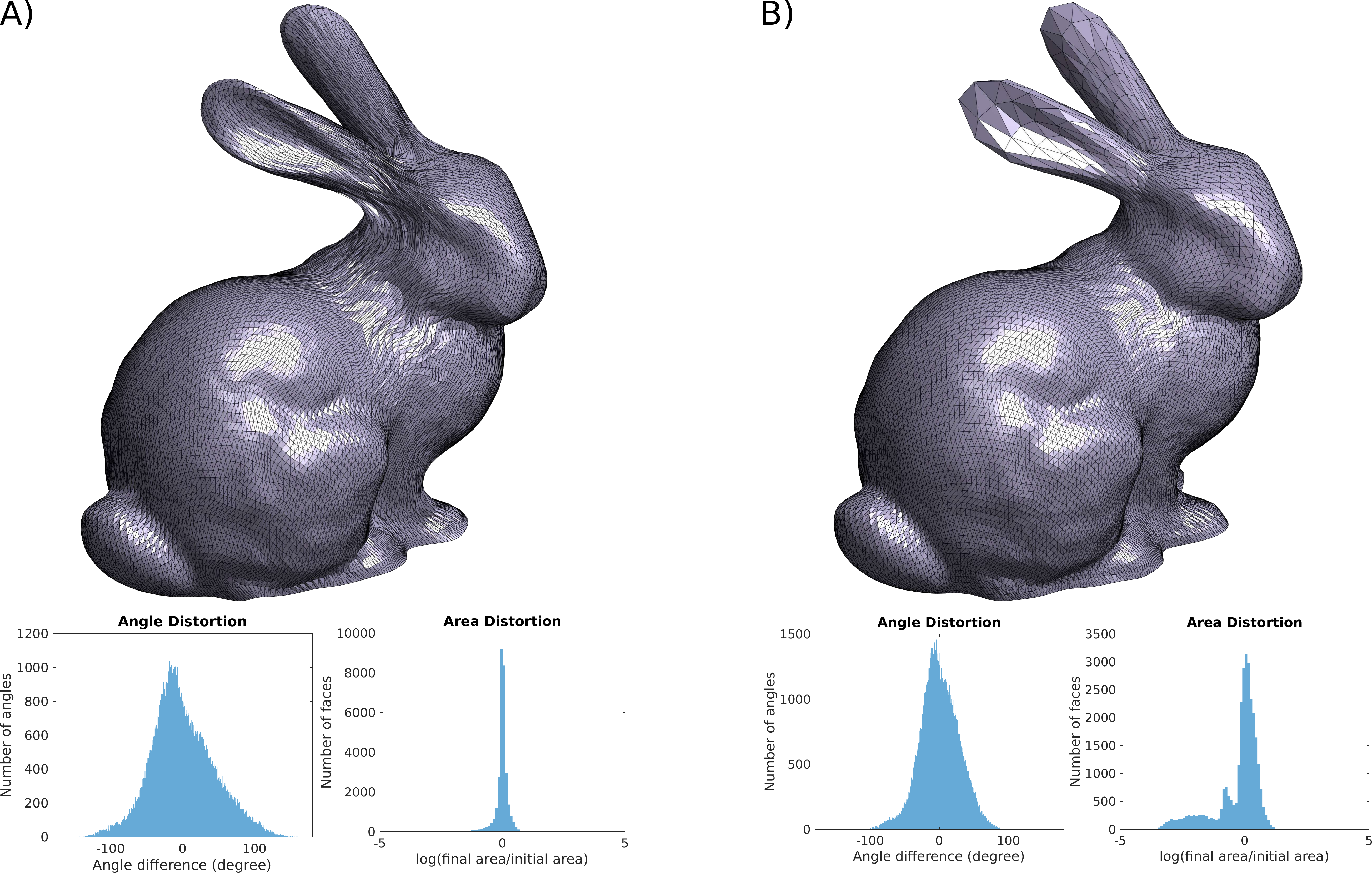}
    \caption{Reconstruction of the Stanford bunny with $n_{max} = 70$ and constituted of $12,546$ vertices (each). The parameterization domain constitutes a prolate hemispheroid with a size of $a = 1, \text{ and } c = 1.0925$. Below each surface, are two histograms for the angle and area distortions. (A) was reconstructed using the proposed area-preserving algorithm AP-HSOH; the reconstruction error, in this case, A-RMSE = $0.000731$. (B) was reconstructed using the balanced mapping with $\alpha = 0$, $\beta = 0.35$, and $\gamma = 0.65$; the reconstruction error, in this case, A-RMSE = $0.000576$. For this example, we assumed $\epsilon_{\eta} = \pi/100$ for the correct reconstruction of the open edge.}
    \label{fig:bunny_rec}
\end{figure}

The Stanford bunny, with an open base, is a challenging surface that usually causes reconstruction issues with harmonic approaches. In our DHA \cite{shaqfa2023disk} work, we failed to correctly reconstruct the ears of the bunny that comprised a relatively small patch on the unit disk after parameterization. Notwithstanding, in this paper, we successfully analyzed and reconstructed the surface using the proposed AP-HSOH approach. Figure \ref{fig:bunny_rec}A shows the reconstruction results using the AP-HSOH approach. The AP-HSOH approach successfully reconstructed the global features of the mesh. However, the texture of the reconstructed surface on the back of the bunny and the ears' connectivity to the head is noisy.

\par
To obtain a better reconstruction, we computed a balanced mapping that minimizes the R-RMSE reconstruction error (B-HSOH). To find the optimal Beltrami weights, we used a heuristic optimization approach to minimize the reconstruction error A-RMSE (our objective function). To limit the needed computational times of our objective, we set $n_{max} = 10$ to measure the reconstruction error during the optimization process; we here fixed the hemispheroidal radius $c$ to the depth of the bounding box. This approach follows our proposal in the SCH paper \cite{shaqfa2021spherical} to find the optimal cap size via the heuristic algorithm in \cite{Shaqfa2021_PSS}. The optimal Beltrami weights were $\beta = 0.3818$ (conformal weight) and $\gamma = 0.6182$ (area-preserving weight), however with the manual inspection we found that $\beta = 0.35$ and $\gamma = 0.65$ gave slightly better results. With the latter composition, the overall reconstruction error dropped from A-RMSE = $0.000731$ with the AP-HSOH to $0.000576$ using the B-HSOH (see Fig.~\ref{fig:bunny_rec}B). Also, notice the change in the histogram distribution for the angle and area distortions in both approaches. As we mixed the AP mapping with the conformal one in this optimal reconstruction, we obtained a smooth surface texture, however, the ears' details were slightly compromised.

\subsubsection{Relationship between mapping and reconstruction accuracy}
\par
In the former example, we saw how the optimal distortion metrics do not necessarily mean optimal reconstruction. Here, we demonstrate the relationship between mapping and reconstruction accuracy. For this, we use the 3D face surface that was formerly used as a benchmark in our works Shaqfa et al. (2021) \cite{shaqfa2021spherical} and (2024) \cite{shaqfa2023disk}.
\par
\begin{figure}[!t]
    \centering
    \includegraphics[width=\linewidth]{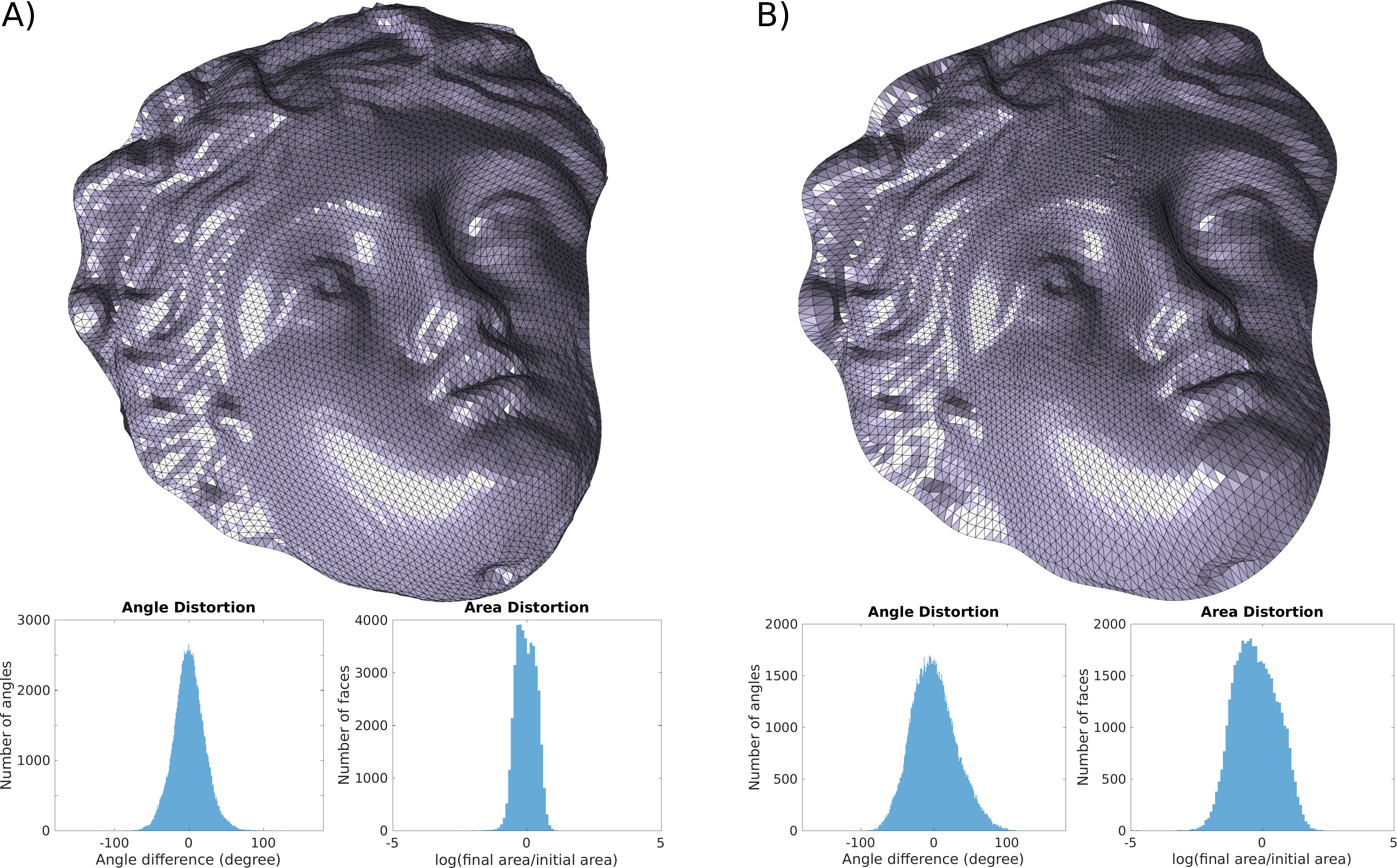}
    \caption{Reconstruction of the 3D face benchmark surface with $n_{max} = 75$ where the input surface constituted of $21,280$ vertices. The parameterization domain constitutes an oblate hemispheroid with a size of $a = 1, \text{ and } c = 0.3938$. The insets underneath show both the angle and area distortions associated with each reconstruction. (A) shows the reconstruction results via the balanced mapping with optimal $\alpha = 0.3440$, $\beta = 0.0$, and $\gamma = 0.6560$ before the harmonic expansion (B-HSOH) of the surface with a reconstruction accuracy A-RMSE = $0.000389$. (B) We used only Tutte mapping for the parameterization before the harmonic expansion step (T-HSOH), the RMSE of Hausdorff distance was $0.000959$. In this example we used $\epsilon_{\eta} = \pi/160$.}
    \label{fig:face_rec}
\end{figure}

For this example, we used balanced mapping to decompose and reconstruct the surface (B-HSOH). Like the Stanford bunny example, the optimal Beltrami weights obtained were $\alpha = 0.344$ (Tutte weight) and $\gamma = 0.656$ (area-preserving weight). The B-HSOH reconstruction results are shown in Fig.~\ref{fig:face_rec}A. The A-RMSE of the B-HSOH approach was $0.000389$, while the DHA \cite{shaqfa2023disk} approach was slightly better with A-RMSE = $0.000202$; both surfaces were visually matching. For the SCH \cite{shaqfa2021spherical} approach we only could construct the basis up $n_{max} = 40$, so we did not consider it here for comparison.
\par
Figure~\ref{fig:face_rec}B shows the T-HSOH approach where the surface shows some textural noise. Underneath each reconstruction result, the corresponding angle and area distortion metrics are shown. Judging only by the parameterization metrics, one would think that T-HSOH reconstruction will not be as good as shown. This is because these metrics are not directly correlated with the decomposition-reconstruction problem.

\begin{figure}[t!]
    \centering
    \includegraphics[width=1.0\linewidth]{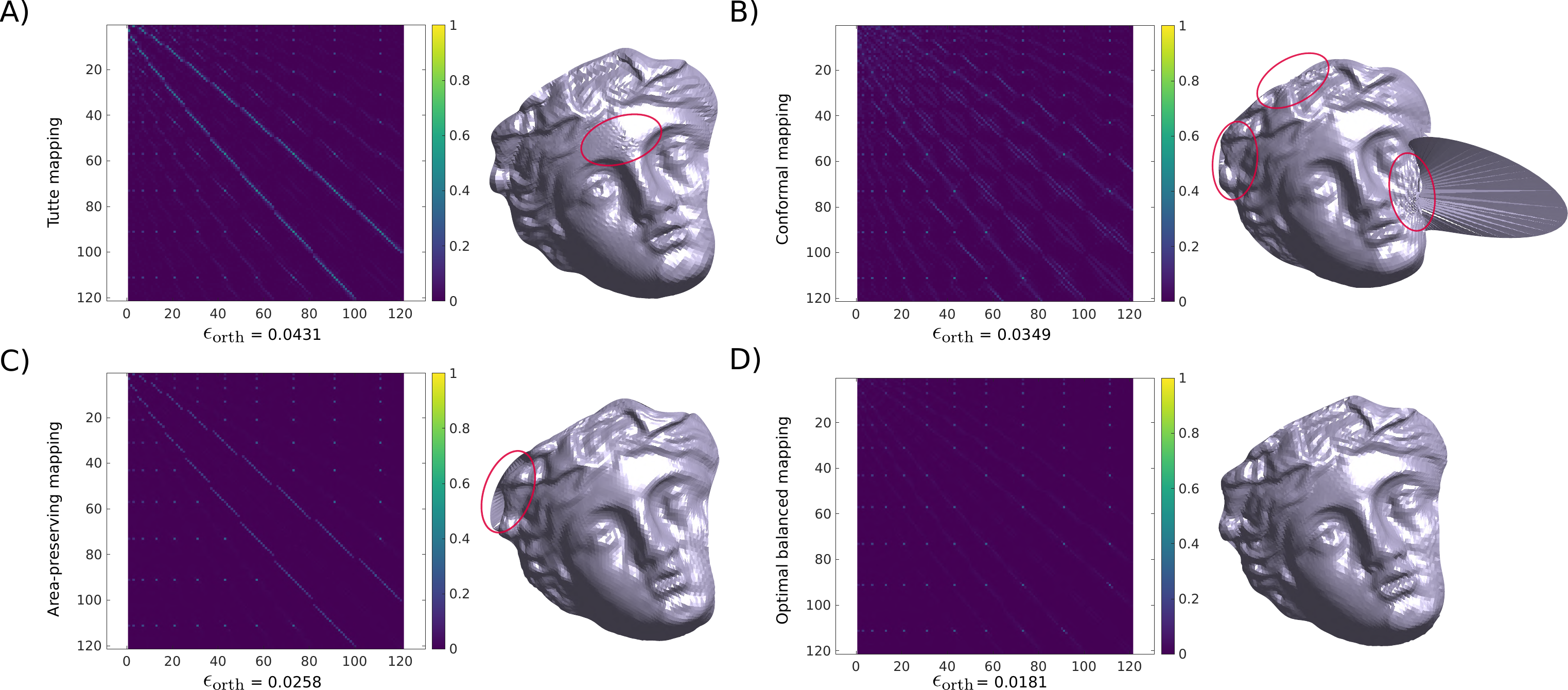}
    \caption{The effect of orthogonality error of the basis function on the reconstruction accuracy for the Tutte approach T-HSOH (A), the conformal approach C-HSOH (B), the area-preserving approach AP-HSOH (C), and the balanced approach B-HSOH (D). The left column shows the absolute difference between the orthogonality of the HSOH basis defined on the obtained map and the basis on a uniform map for $n_{max} = 10$ (121 bases). The right column shows the corresponding reconstruction associated with each map. The annotated $\epsilon_{\text{orth}}$ is the total obtained error between the two sets of basis functions. In these examples we used $\epsilon_{\eta} = \pi/160$.}
    \label{fig:ortho_rec_error}
\end{figure}

\par
To understand this behavior, we constructed the orthogonality matrix for each of T-HSOH, C-HSOH, AP-HSOH, and B-HSOH (with $\alpha = 0.3440$ and $\gamma = 0.6560$). Each one of the above-mentioned mappings has a different distribution of the vertices onto the target hemispheroidal domain. By construction, we assume that the basis functions evaluated at the target domain are always orthonormal. The latter statement is satisfied onto a hemispheroid with uniformly distributed vertices. Figure~\ref{fig:ortho_rec_error} shows the orthogonality errors associated with different mappings due to the maps' misalignment with the mathematical grid. The insets show the absolute difference between the basis functions evaluated over the obtained mapping and the ones evaluated over a uniformly generated grid. The orthogonality error, denoted by $\epsilon_{\text{orth}}$, is measured as the mean of this absolute difference between the basis functions. As can be seen, both T-HSOH and C-HSOH give a relatively large orthogonality error, and the reconstruction results contain notable artifacts. AP-HSOH gives a smaller orthogonality error, while minor artifacts can still be observed. The minimal orthogonality error, however, was obtained by B-HSOH as it corresponds to the best reconstruction accuracy (A-RMSE). It should be noted that the balanced approach requires further optimizing the Beltrami weights and is computationally expensive. This observation is similar to the one made by Shaqfa and van Rees~\cite{shaqfa2024spheroid}.

\subsubsection{Further optimizing the hemispheroidal size for the reconstruction}
\par
In the former benchmarks, we followed the registration approach in Section \ref{fig:surface_registration} to determine the hemispheroidal radius $c$. In particular, we only discussed the optimal reconstruction by manipulating the Beltrami weights to find the optimal composition of mappings, where the objective function required surface reconstruction and directly comparing the output surface with the input one. Here, we investigate whether the proposed registration approach is justifiable and if one can further change the hemispheroidal size to optimize the reconstruction result. We here assume the mapping to be area-preserving and we only solve to find the optimal hemispheroidal radius $c$.
\par
Reconstructing the whole surface and comparing it in each iteration is computationally expensive and requires a lot of resources. Moreover, as we assumed $n_{max} = 10$ in the previous examples, that objective function might not reflect the state of higher harmonics error. For this, we here use another indicative objective function that is computationally inexpensive and reflects the state of orthogonality of the basis functions on the mapped hemispheroidal image. From Eq.~\eqref{eqn:general_decomp_final}, we start by normalizing the basis functions $\forall S^n_m \in \bold{S}$, such that $S^n_m \cdot S^{n, T}_m = 1$ (where the $(\cdot)^T$ denotes the transpose). Then, the new objective function is set to minimize the mean value $\bar{\epsilon}_{\text{orth}}$ of the matrix $\bold{S} \bold{S}^T$. Such objective function contains the overall orthogonality error that is proportional to the reconstruction quality.
\par
\begin{figure}[t]
    \centering
    \includegraphics[width = \linewidth]{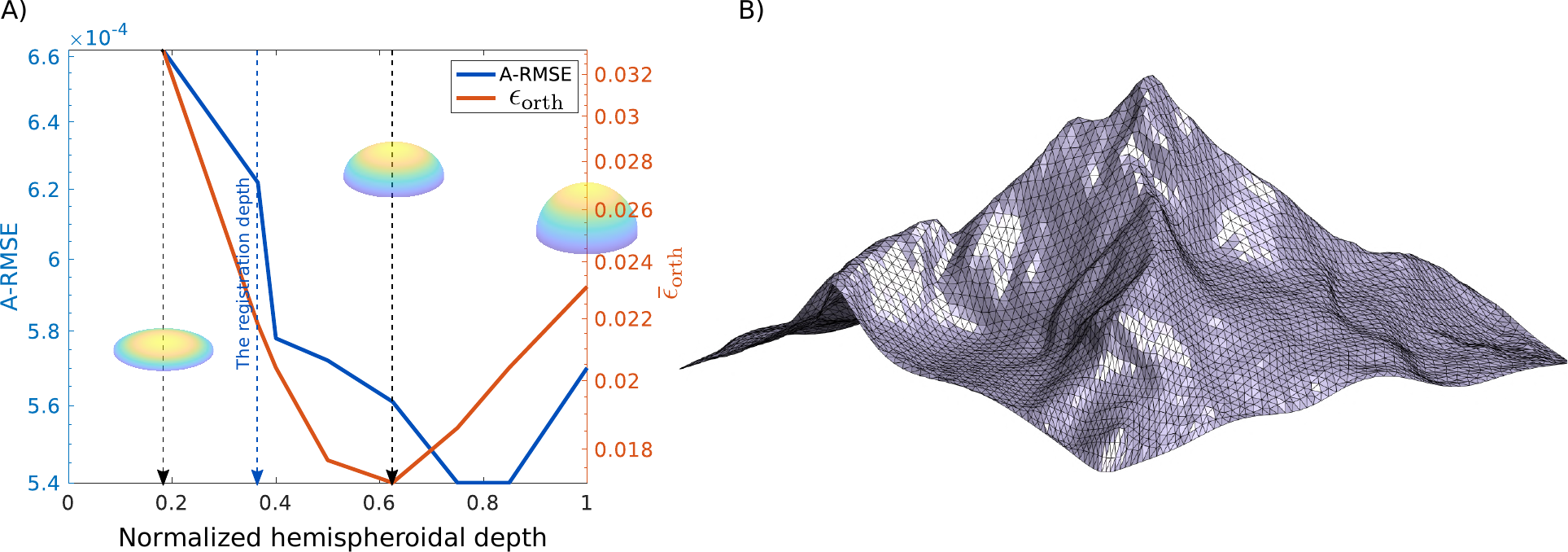}
    \caption{The hemispheroidal size optimization of the Matterhorn benchmark \cite{shaqfa2023disk}. (A) shows the reconstruction error (A-RMSE) vs. normalized radius $c$ of the hemispheroid. The same inset also shows the orthogonality error ($\bar{\epsilon}_{\text{orth}}$). (B) shows the reconstructed surface using AP-HSOH approach with $n_{max} = 50$ and at a normalized radius of $c = 0.625$.}
    \label{fig:matterhorn_size_opt}
\end{figure}

Figure \ref{fig:matterhorn_size_opt}A shows the relationship between the normalized radius $c$ and the reconstruction error A-RMSE as well as the orthogonality one $\bar{\epsilon}_{\text{orth}}$. As can be seen, the general error trends are matching. The minimum points for each of the shown error metrics are not coinciding but they are practically the same as the reconstruction is visually matching. Also, as annotated in Fig.~\ref{fig:matterhorn_size_opt}A, the registration value from the proposed approach in Fig.~\ref{fig:surface_registration} is not far from the optimal value; thus, it can be directly used for most surfaces. Figure \ref{fig:matterhorn_size_opt}B shows the reconstructed benchmark at the radius that corresponds to the minimum value of $\bar{\epsilon}_{\text{orth}}$. In general, the A-RMSE is a more direct measure, however, given the current results and the computational savings we can recommend minimizing $\bar{\epsilon}_{\text{orth}}$ instead. The reconstruction via the DHA approach (see Shaqfa et al.~\cite{shaqfa2023disk}) is slightly better than the AP-HSOH one with an A-RMSE = $0.000491$ (DHA) and no visual discrepancies between both approaches. Altogether, the above experiment suggests that the proposed registration approach in Fig.~\ref{fig:surface_registration} is a simple but effective approach for determining a suitable shape of the hemispheroidal domain for an accurate hemispheroidal harmonic reconstruction.

\section{Conclusion} \label{sect:conclusion}
In this work, we have proposed four methods for the hemispheroidal parameterization of simply connected open surfaces, namely, Tutte mappings, conformal mappings, area-preserving mappings, and balanced mappings. In particular, the use of hemispheroids allows us to achieve a lower geometric distortion when parameterizing any given simply connected open surface. Then, by combining hemispheroidal parameterization and hemispheroidal harmonics, we have developed a new framework for the shape representation and harmonic decomposition of simply connected open surfaces. Our experimental results have demonstrated the effectiveness of the proposed approach for handling a large variety of surfaces with different geometries.
\par
As shown in our analysis, the area-preserving mapping was found to have a consistent reconstruction performance in most of the tested cases. Specifically, suppose the vertex distribution on the input surface mesh is sufficiently uniform. In that case, the area-preserving mapping will normally provide a largely uniform mesh on the target hemispheroid with good orthonormality, thereby giving good reconstruction results. More generally, the best reconstruction results can be achieved by further optimizing the weighting parameters in the balanced mapping approach. 
\par
Note that the parameterization and harmonic decomposition methods presented in this work are limited to simply connected open surfaces but not surfaces with other topologies. In the future, we plan to extend our work for the shape representation and analysis of multiply connected surfaces.

\bibliographystyle{ieeetr}
\bibliography{reference.bib}

\end{document}